\documentclass[12pt]{article}%
\usepackage{amsfonts}
\usepackage{amssymb}
\usepackage{graphicx}
\usepackage{eurosym}

\usepackage{amsmath}%
\title{Analyzing Risky Choices: Q-Learning for Deal-No Deal}
\author{Laszlo Korsos and Nicholas G. Polson\footnote{Korsos is at the Booth School of Business, 
University of Chicago, 5807 S. Woodlawn Avenue Chicago, IL 60637. Polson is Professor
of Econometrics and Statistics at  the Booth School of Business, 
University of Chicago, 5807 S. Woodlawn Avenue Chicago, IL 60637. Emails {\tt [lkorsos,ngp]@chicagobooth.edu}}}

\begin{document}
\maketitle

\begin{abstract}
\noindent We derive an optimal strategy in the popular \textit{Deal or No Deal} game show.  
Q-learning quantifies the continuation value inherent in sequential decision
making and we use this to analyze contestants risky choices. Given their choices and 
optimal strategy, we invert to find implied bounds on   
their levels of risk aversion. In risky decision making, previous empirical evidence has suggested that
past outcomes affect future choices and that contestants have time-varying risk aversion. 
We demonstrate that the strategies of two players (Suzanne and Frank) from the European version of the game 
are consistent with constant risk aversion levels except for their last risk-seeking choice.

\end{abstract}

\section{Introduction}
Ever since the introduction of the popular television show \textit{Deal or No Deal}, many authors have analyzed aspects of the game. \textit{Deal or No Deal} provides  an `experiment' with large stakes and a relatively simple probabilistic structure. Using Q-learning (Watkins, 1989, Whittle, 1982, Putterman, 1984, Polson and Sorensen, 2011) we address the question of optimal strategy. 
Our solution technique provides 
a dynamic maximum expected utility solution (Ramsey, 1926, de Finetti, 1937,
von Neumann and Mortgensen, 1944). 
Q-learning is a popular reinforcement learning technique for calculating continuation values for sequential
decision making.

In \textit{Deal or No Deal} contestants are presented with a simple decision of whether to
take a Banker's offer or to continue in the game.
Hence the value realized from deciding to decline the banker's offer - `No Deal' - has a continuation value, a stylized fact of 
sequential decision making.  Previous empirical evidence has suggested that
past outcomes affect future choices and that contestants have time-varying risk aversion.
One key feature of continuation values is that they can appear to lead to the same effect -- namely
current actions appearing to exhibit time-varying risk aversion --  when in fact it is no more than an optimal action to exercise the continuation value of the game. 
To illustrate these effects, 
we consider a simple scenario with logarithmic
utility. This also illustrates a theoretical \emph{rule of thumb} in \textit{Deal or No Deal}: one should generally continue
as long as there are two big prizes left.

We analyze data from two players, Frank and Suzanne, from the 
European version of the game show (Post et al, 2008). Tables 1 and 2 provide the contestants choices.
For example, in round seven, after several unlucky picks, Frank opened the briefcase
with the last remaining large prize (\euro $ 500\mathord{,}000 $) and saw his expected prize tumble
from \euro $ 102\mathord{,}006 $ to \euro $2\mathord{,}508$. The banker offered him \euro $2\mathord{,}400$ but Frank rejected the offer
and continued to play. He finally ended up with a briefcase with only \euro $10$.
In round nine, he even rejected a certain \euro $6\mathord{,}000$ in favor of a $50/50$ gamble of \euro $10$ or \euro $10\mathord{,}000$ -- clearly exhibiting risk seeking behavior.

In contrast, Suzanne was a ``lucky'' player. In round nine see faced a $50/50$ gamble of \euro $100\mathord{,}000$
or \euro $150\mathord{,}000$ (two of the three largest prizes in the German edition). While she was
hesitant in the earlier rounds, she rejected the banker's offer of \euro $125\mathord{,}000$ -- the expected payoff -- and finally
won the \euro $150\mathord{,}000$ prize. Our analysis will track their choices and infer bounds on their levels of risk aversion
at each stage. We provide a separate analysis of the last stage of the game, as both players exhibit
risk seeking behavior here.
Other authors claim declining risk aversion of individuals after earlier expectations have been shattered by unfavorable outcomes. Our approach
shows that the continuation value is in fact high and therefore risk aversion of contestants need not decline in the aforementioned case.

The rest of the paper is outlined as follows.  In Section 2 we discuss the Q-Learning technique and how it is applied to the \textit{Deal or No Deal} game, including log and power utility examples.  Section 3 solves for optimal strategy using
Q-learning. Given an optimal strategy and the contestants empirical choices, we can then infer bounds on their risk aversion.
We also analyze their terminal risk-seeking choices.
Finally, Section 4 concludes.

\section{Q-Learning and Deal-No Deal}

\subsection{\textit{Deal or No Deal}}

In the game \textit{Deal or No Deal}, players are presented with a choice of briefcases which hold distinct monetary prize values.  Initially, players are asked to select one case, which will be referred to as `their case' and remain unopened.  Then, they proceed by choosing, and thus eliminating a predetermined number of cases each round.  The values of these eliminated cases are shown to the player.  At the end of each set of case eliminations, an entity referred to as `The Banker' then presents the player with a monetary offer in exchange for their case.  At this point, the player is given two options: either take the Banker's offer (i.e. Deal) 
or continue eliminating cases (i.e. No Deal).  The game continues until an offer is accepted or there is only one case remaining.

From this setup, we can decompose the game into a set of states.  Let the set $V$ contain all the possible prize values in the initial suitcases.  
Let the set $S$ consist of all possible combinations of prize values from the given set of possible prize values $V$.  Let, $s\in S$ be one of the sets of possible remaining suitcase prize values.
At each state $s$, the player is given a set of possible actions $A$, defined as follows:
$$
A = \left\{ 0 ,1 \right \} = \left \{ \text{`No Deal'} , \text{`Deal'} \right\}
$$
After performing an action $a\in A$, the player then experiences a transition from state to state.  
We define the payoff to the player as a state-action map onto the space of real numbers:
$ Q:S \times A \to \mathbb{R} $. 
The goal is simply to maximize the discounted resulting value of this mapping.  Q-Learning allows us to maximize utility over a set of possible decisions, providing a map of the optimal path of actions.

\subsection{Optimal Strategy: Q-Learning}

In this subsection, we describe the basic ingredients of Q-learning. 
First, we let the agent's immediate utility, taking the `Deal', be denoted by $ u( s_t, a_t )$. The Bellman
principle of optimality states that the optimal solution path $ a^\star ( s )$ is the solution to
the Bellman equation defined by a value function $V(s)$ that satisfies
$$
V ( s ) = \max_{ a^\prime } \; \left \{ u( s , a^\prime ) + \sum_{ s^\prime }
 V \left ( s ^\prime \right ) p ( s^\prime | s,a^\prime ) \right \}
$$
Here $ p ( s^\prime | s,a^\prime )$ is the transition matrix throughout states given action
$a^\prime$. In \textit{Deal or No Deal}, actions cannot affect the state's evolution, so we write $ p ( s^\prime | s ) $.  
Rather than directly computing the value function, we instead calculate the matrix of $Q$-values.
They are defined as the total expected utility gained by choosing a current action $a$ and following
the optimal path thereafter.

An optimal policy must satisfy Bellman's principle of optimality:
\emph{that an optimal path has the property that whatever the initial conditions and control variables (choices) over some initial period, the control (or decision variables) chosen over the remaining period must be optimal for the remaining problem, with the state resulting from the early decisions taken to be the initial condition.}

To solve for this, we need a transition matrix for probabilities, a utility function and a banker's
valuation function. To fix notation, let $s\in S$ denote the current state of the system and $a\in A$ an action. Define the $Q$-value, $Q_t(s,a)$, at time $t$ by the value of using action $a$ today and then proceeding optimally in the future.
The Bellman equation for $Q$-values becomes:
$$
Q_{t} ( s , a) = u( s , a  ) + \sum_{ s^\star\in S^\star} P( s^\star | s ,a ) \max_{ a\in A } Q_{t+1} ( s^\star , a )
$$
where $S^\star\subset S$ is the set of all possible next period states given the current action $a$, namely
$S^\star=\{s^\star:s^\star\in\mathcal{P}(s)\cap|s^\star|=|s|-1\} $. The 
value function and optimal action are then simply given by:
$$ 
V(s) = \max_{a\in A} Q ( s , a ) \; \; {\rm and} \; \;  a^\star (s) = {\rm arg}\max_{a\in A} \; Q ( s , a ) 
$$ 
In our discrete setting, we can directly find the $Q$-values without resorting to
the simple stochastic approximation algorithms given in Watkins (1989) and Watkins and Dayan (1992).  We now define the ingredients to solve the problem:

\begin{description}
\item[Transition Matrix.]

Since there is equal probability that a player chooses any of the existing prize values, we define the transition probabilities as follows:
$$
P(s^\star|s,a=1) = \frac{1}{|S^\star|}=\frac{1}{|s|}
$$
For example, when you have three prizes left, with $s$ the current state 
$$
S^\star = \{ {\rm all \; subsets \; of \; two \; prizes} \} \; \; {\rm and} \; \; P( s^\star | s, a =1) = \frac{1}{3}
$$
where the transition matrix is uniform to the next state.

There is no continuation value for taking the Deal.  If action $a=0$ is chosen, then the value realized by the player is either the utility of the offer presented by the banker or the utility of the value of the player's case (if no other cases remain).

\item[Banker's Function $B(s)$.]
There are a number of different choices for modeling the banker's function.  In the live TV show one only sees the current banker offer and of course, to address the issue of optimal policy and the continuation value we need to know what they would offer in future states of the world.  One popular choice is expected value: let $v \in S$ be a possible prize, then 
$$
B( s ) = \bar{s} \equiv \frac{1}{|s|}\sum_{v\in s}v
$$
where $s$ is the set of the remaining prize values. Another choice is from the on-line version
of the game where the website ({\tt www.nbc.com/DealOrNoDeal}) uses the following criteria:
let \emph{big} and \emph{small} denote the biggest and smallest prizes left on the board, respectively. Then, the
banker offer is given by 

\begin{itemize}
\item With 3 prizes left:   $ B( s) = 0.305 \cdot {\rm big} + 0.5 \cdot {\rm small}$

\item With 2 prizes left:   $B( s) = 0.355 \cdot {\rm big} + 0.5 \cdot {\rm small}$.
\end{itemize}

This is not the case in the TV show as the Banker has some discretion on the offer.
Empirically, it almost strictly holds that: $B(s)<\bar{s}$ -- the expected value of the remaining prizes.

\item[Utility.]

The utility of the next state depends on the contestant's value for money and the bidding function $B(s)$ of the banker.  For example, 
in the case of CRRA power utility
$$
u( B ( s ) ) = \frac{ B ( s )^{1-\gamma} -1 }{1 - \gamma }
$$
with log-utility 
$u(B(s)) = \ln(B(s)) $ a special case. 
A more flexible choice is the exponential-power utility function 
$$
 u ( B(s) ) = \alpha^{-1} \left ( 1 - \exp \left ( - \alpha ( W + B(s) )^{1- \gamma} \right ) \right ) 
$$ 
where $W$ is current wealth.
For the purpose of our analysis, we focus on the natural logarithm and CRRA cases. 
\end{description}

\subsection{Illustrative Example: Log-Utility}

To show that the continuation value can be large we consider an example
where there are three prizes left including two large ones, $s = \{ 750 , 500 , 25 \} $.
Let us take an example where the contestant is risk averse with log-utility:
$u(x) = \ln x $. This is equivalent to the well-known Kelly (1956) criterion.
The contestant would be indifferent to a coin toss that doubled or halved their wealth.

For a base case analysis, suppose that the Banker's offers are determined by
the expected value of the prizes left in the set $s$.  With log-utility this will look
like a good deal in a one-shot version of the game. For this example, the utility of the offer is
$$
u( B( s = \{ 750 , 500 , 25 \}) ) = \ln ( 1275/3  ) = 6.052
$$
Taking the deal leads to a utility $ Q_t ( s , a= 0 ) = 6.052 $.

However, we have to compare this to the continuation problem (`No Deal').  The set of future possible states is
$S^\star=\{ s_1^\star , s_2^\star , s_3^\star  \} $
where
$$
s_1^\star  = \{750,500\} \; , \; 
s_2^\star = \{750,25\} \; , \; 
s_3^\star = \{500,25\}
$$
As the banker offers the expected value, if the contestant picks `No Deal' we will have offers of $625$, $387.5$, and $137.5$, respectively.  This gives the following $Q$-value calculation:
\begin{align}
Q_{t} ( s , a=1) & = \sum_{ s^\star\in S^\star } P( s^\star | s ,a =1) \max_{ a\in A } Q_{t+1} ( s^\star , a ) \nonumber\\
& = \frac{1}{3} \left (  \ln (625) + \ln (387.5) + \ln (262.5) \right ) = 5.989 \nonumber
\end{align}
with immediate utility given by $ u(s,a)=0$.

Therefore, as $ Q_{t} ( s , a=1)=5.989 <  6.052 = Q_{t} ( s , a=0)$ the optimal 
action for the player at time $t$ is $ a^\star = 0 $, `Deal'.  
We see that the continuation value was not large enough to overcome the generous (expected value) offer by the banker.

This analysis also illustrates the rule-of-thumb that most players should proceed as long as there are two large prizes
left as the continuation value is high.

\subsection{Sensitivity Analysis - Different Banker's Function}

The magnitude of the continuation value is related to the Banker's bidding function. 
If we now use the Banker's bidding function provided by the on-line game, defined by:
$$
B( s) = 0.355 \cdot {\rm big} + 0.5 \cdot {\rm small} \;\; {\rm (with \; two \; prizes \; remaining)} 
$$
Hence, we can evaluate the banker's function at each of the $s^*$ next period states:
\begin{align}
B( s_1^\star = \{750,500 \}) &= 516.25\nonumber\\
B( s_2^\star = \{ 750,25 \}) &= 278.75\nonumber\\
B( s_3^\star = \{ 500,25 \}) &= 190\nonumber
\end{align}
Using these, let us consider the optimal action with 2 prize values left for the player.  To do so, we calculate the following $Q$-values:
\begin{align}
Q_{t+1} ( s_1^\star , a=1) &= \frac{1}{2} \left \{  \ln (750) + \ln (500) \right \} = 6.415\nonumber\\
Q_{t+1} ( s_1^\star , a=0) &= \ln \left ( 516.25 \right ) = 6.246\nonumber
\end{align}
Since $Q_{t+1} ( s_1^\star , a=1)>Q_{t+1} ( s_1^\star , a=0)$, the future optimal policy is `No Deal' under $s_1^\star$.
\begin{align}
Q_{t+1} ( s_2^\star , a=1) &= \frac{1}{2} \left \{  \ln (750) + \ln (25) \right \} = 4.9194\nonumber\\
Q_{t+1} ( s_2^\star , a=0) &= \ln \left ( 278.75 \right ) = 5.63\nonumber
\end{align}
Since $Q_{t+1} ( s_2^\star , a=1)<Q_{t+1} ( s_2^\star , a=0)$, the future optimal policy is `Deal' under $s_2^\star$.
\begin{align}
Q_{t+1} ( s_3^\star , a=1) &= \frac{1}{2} \left \{  \ln (500) + \ln (25) \right \} = 4.716\nonumber\\
Q_{t+1} ( s_3^\star , a=0) &= \ln \left ( 190 \right ) = 5.247\nonumber
\end{align}
Since $Q_{t+1} ( s_3^\star , a=1)<Q_{t+1} ( s_3^\star , a=0)$, the future optimal policy is `Deal' under $s_3^\star$.

Now, solving for $Q$-values at the previous step gives
\begin{align}
Q_{t} ( s , a=1) & = \sum_{ s^\star\in S^\star } P( s^\star | s ,a =1) \max_{ a\in A } Q_{t+1} ( s^\star , a ) \nonumber\\
& = \frac{1}{3} \left (  6.415 + 5.63 + 5.247 \right ) = 5.764\nonumber
\end{align}
with a monetary equivalent of $ \exp(5.764  ) = 318.62 $.  This is the continuation value (or `No Deal' decision value) at the current time period $t$.  We can compare this with the `Deal' decision value:
\begin{align}
Q_{t} ( s , a=0) &= u( B( s = \{ 750 , 500 , 25 \}) )\nonumber\\
&= \ln \left ( 0.305 \cdot 750 + 0.5 \cdot 25 \right )\nonumber\\
&= \ln ( 241.25 ) = 5.48\nonumber
\end{align}
The Banker then offers the contestant a monetary value of $241.25$.  Now, comparing the $Q$-values of the decisions
gives the optimal action of $a^\star = 1$, `No Deal' as
$$
Q_{t} ( s , a=1)= 5.7079  > 5.48 = Q_{t} ( s , a=0)
$$
We get this result because the continuation value is large. Essentially we are considering the difference between  $\$241$ compared to $\$319$, or a $33$\% premium.

To extend this analysis to include higher levels of risk aversion we consider can power utility:
$$
u(x) = \frac{ x^{1-\gamma} -1 }{1 - \gamma }
$$
where $ \gamma = - u^{\prime \prime} (x)/u(x) $ is a local measure of risk aversion, first
introduced by de Finetti (1952).  We note that as $\gamma$ tends to 1, the utility function above tends toward the aforementioned logarithmic utility function.  Therefore, the logarithmic utility function is just a special case of the more general power utility function.

With the same three prize values remaining:
$s = \{ 750 , 500 , 25 \}  $ and an the expected value criterion for the banker's function, we get
utilities:
$$
u( B( s = \{ 750 , 500 , 25 \}) ) = \frac{ (1275/3)^{1-\gamma} -1 }{1 - \gamma }
$$
Therefore, the utility of `Deal' is $ Q_t ( s , a= 0 ) = \frac{ (425)^{1-\gamma} -1 }{1 - \gamma } $.

As before, we consider the continuation problem (`No Deal') of the set
$S^\star=\{ s_1^\star , s_2^\star , s_3^\star  \} $ with
expected value banker offers leading to the values $625$, $387.5$, and $137.5$, respectively.  
Performing the $Q$-value calculation gives:
\begin{align}
Q_{t} ( s , a=1) & = \sum_{ s^\star\in S^\star } P( s^\star | s ,a =1) \max_{ a\in A } Q_{t+1} ( s^\star , a ) \nonumber\\
& = \frac{1}{3} \left (  \frac{ (625)^{1-\gamma} -1 }{1 - \gamma } + \frac{ (387.5)^{1-\gamma} -1 }{1 - \gamma } + \frac{ (262.5)^{1-\gamma} -1 }{1 - \gamma } \right ) \nonumber\\
& = \frac{(625)^{1-\gamma}+(387.5)^{1-\gamma}+(262.5)^{1-\gamma}-3}{3(1-\gamma)}\nonumber
\end{align}
with immediate utility $ u( s,a ) = 0 $.

Hence, as the inequality $ Q_{t} ( s , a=1) < Q_{t} ( s , a=0)$ holds
 for all levels of risk aversion $0<\gamma<\infty$, the optimal action is $ a^\star = 0 $, `Deal'.  Once again, we can see that the continuation value was not large enough to overcome the generous (expected value) offer by the banker.

\subsection{Sensitivity Analysis - Different Banker's Function}

Now, using the Banker's bidding function from the website, we once again have:
$$
B( s) = 0.355 \cdot {\rm big} + 0.5 \cdot {\rm small} \;\; {\rm (with \; two \; prizes \; remaining)} 
$$
From the previous section, we found that $B(s_1^\star)=516.25$, $B(s_2^\star)=278.75$, and $B(s_3^\star)=190$.

We now consider the optimal action with 2 prize values left for the player.  To do so, we calculate the following $Q$-values:
\begin{align}
Q_{t+1} ( s_1^\star , a=1) &= \frac{1}{2} \left ( \frac{ 750^{1-\gamma} -1 }{1 - \gamma }  + \frac{ 500^{1-\gamma} -1 }{1 - \gamma } \right ) = \frac{750^{1-\gamma}+500^{1-\gamma}-2}{2(1-\gamma)}\nonumber\\
Q_{t+1} ( s_1^\star , a=0) &= \frac{ 516.25^{1-\gamma} -1 }{1 - \gamma }\nonumber
\end{align}
Since $Q_{t+1} ( s_1^\star , a=1)>Q_{t+1} ( s_1^\star , a=0)$ for all levels of risk aversion $0<\gamma<\infty$, the future optimal policy is `No Deal' under $s_1^\star$.
\begin{align}
Q_{t+1} ( s_2^\star , a=1) &= \frac{1}{2} \left (  \frac{ 750^{1-\gamma} -1 }{1 - \gamma } + \frac{ 25^{1-\gamma} -1 }{1 - \gamma } \right ) = \frac{750^{1-\gamma}+25^{1-\gamma}-2}{2(1-\gamma)}\nonumber\\
Q_{t+1} ( s_2^\star , a=0) &= \frac{ 278.75^{1-\gamma} -1 }{1 - \gamma }\nonumber
\end{align}
Here, for $0<\gamma<0.5602$, we find that $Q_{t+1} ( s_2^\star , a=1)>Q_{t+1} ( s_2^\star , a=0)$  (so the optimal future policy is `No Deal' under $s_2^\star$).  On the other hand, for $0.5602<\gamma<\infty$, we find that $Q_{t+1} ( s_2^\star , a=1)<Q_{t+1} ( s_2^\star , a=0)$ (so the future optimal policy is `Deal' under $s_2^\star$).
\begin{align}
Q_{t+1} ( s_3^\star , a=1) &= \frac{1}{2} \left ( \frac{ 500^{1-\gamma} -1 }{1 - \gamma } + \frac{ 25^{1-\gamma} -1 }{1 - \gamma } \right ) = \frac{500^{1-\gamma}+25^{1-\gamma}-2}{2(1-\gamma)}\nonumber\\
Q_{t+1} ( s_3^\star , a=0) &= \frac{ 190^{1-\gamma} -1 }{1 - \gamma }\nonumber
\end{align}
Here, for $0<\gamma<0.5175$, we find that $Q_{t+1} ( s_3^\star , a=1)>Q_{t+1} ( s_3^\star , a=0)$  (so the optimal future policy is `No Deal' under $s_3^\star$).  On the other hand, for $0.5175<\gamma<\infty$, we find that $Q_{t+1} ( s_3^\star , a=1)<Q_{t+1} ( s_3^\star , a=0)$ (so the future optimal policy is `Deal' under $s_3^\star$).

Solving for the $Q$-values at the previous step gives
\begin{align}
Q_{t} ( s , a=1) & = \sum_{ s^\star\in S^\star } P( s^\star | s ,a =1) \max_{ a\in A } Q_{t+1} ( s^\star , a ) \nonumber\\
& = \left\{
\begin{array}{ll}
\frac{1}{3} \left ( \frac{750^{1-\gamma}+500^{1-\gamma}-2}{2(1-\gamma)} + \frac{750^{1-\gamma}+25^{1-\gamma}-2}{2(1-\gamma)} +  \frac{500^{1-\gamma}+25^{1-\gamma}-2}{2(1-\gamma)}\right ) & \text{if } 0<\gamma<0.5175\\
\frac{1}{3} \left ( \frac{750^{1-\gamma}+500^{1-\gamma}-2}{2(1-\gamma)} + \frac{750^{1-\gamma}+25^{1-\gamma}-2}{2(1-\gamma)} + \frac{ 190^{1-\gamma} -1 }{1 - \gamma }\right ) & \text{if } 0.5175<\gamma<0.5602\\
\frac{1}{3} \left ( \frac{750^{1-\gamma}+500^{1-\gamma}-2}{2(1-\gamma)} + \frac{ 278.75^{1-\gamma} -1 }{1 - \gamma } + \frac{ 190^{1-\gamma} -1 }{1 - \gamma }\right ) & \text{if } 0.5602<\gamma<\infty
\end{array} \right. \nonumber
\end{align}
From this, we can calculate the following monetary equivalents:
\begin{align}
\left((1-\gamma)Q_{t} ( s , a=1)+1\right)^{\frac{1}{1-\gamma}}=
\left\{
\begin{array}{ll}
\left(\frac{750^{1-\gamma}+500^{1-\gamma}+25^{1-\gamma}}{3}\right)^{\frac{1}{1-\gamma}}& \text{if } 0<\gamma<0.5175\\
\left(\frac{2(750^{1-\gamma}+190^{1-\gamma})+500^{1-\gamma}+25^{1-\gamma}}{6}\right)^{\frac{1}{1-\gamma}}& \text{if } 0.5175<\gamma<0.5602\\
\left(\frac{2(278.75^{1-\gamma}+190^{1-\gamma})+750^{1-\gamma}+500^{1-\gamma}}{6}\right)^{\frac{1}{1-\gamma}}& \text{if } 0.5602<\gamma<\infty
\end{array} \right.\nonumber
\end{align}
These are the continuation values (or `No Deal' decision values) at the current time period.  We can compare these with the `Deal' decision value:
\begin{align}
Q_{t} ( s , a=0) &= u( B( s = \{ 750 , 500 , 25 \}) )\nonumber\\
&= \frac{ (0.305 \cdot 750 + 0.5 \cdot 25)^{1-\gamma} -1 }{1 - \gamma }\nonumber\\
&= \frac{ (241.25)^{1-\gamma} -1 }{1 - \gamma }\nonumber
\end{align}
Hence, the Banker offers the contestant a monetary value of $241.25$.  Now, comparing the $Q$-values of the decisions:

\begin{itemize}
\item $Q_{t} ( s , a=1) >  Q_{t} ( s , a=0)$ if $0<\gamma<4.5963 $ with optimal action $a^\star=1$, `No Deal'.
\item $Q_{t} ( s , a=0) >  Q_{t} ( s , a=1)$ if $4.5963<\gamma<\infty$ with optimal action $a^\star=0$, `Deal'.
\end{itemize}

Therefore, we can see that for most risk aversion levels (except for really high values), the optimal action will be to choose `No Deal'.  This is because the continuation value is high relative to the value of the banker's offer.
Once again, for common risk aversion levels, we can approximately reduce the Q-Learning results to a simple rule of thumb.  That is, continue as long as there are two large prizes left.

\section{Analyzing Risky Choices: Suzanne and Frank}

A number of authors have argued that backwards induction appears most relevant in the early rounds of the game and that there is no difference with the myopic rule as the 
Banker's offers (near expected value) lead risk averse players to proceed.
Our approach has shown that there is a significant continuation value at 
the end of the game.  Take for example, a set of three prizes containing two large ones.  Risk averse people will naively choose the action 
`Deal', when if they incorporated the continuation value, they would choose `No Deal'.  Of course, this is sensitive to the banker's offer function.  If risk aversion is very high as manifested by the contestant's utility function, then the continuation value may not always be high.

Post et al (2008) proposes that path-dependence factors 
heavily into the choices of contestants, and in fact the choices can be explained by varying levels of risk aversion as the game progresses favorably or unfavorably for the contestant.  We find that this does not have to necessarily be the case since choosing to turn down a Banker's offer does not have to imply decreasing risk aversion, but only a higher continuation value present by removing the myopia restriction/assumption on the contestants.

As well, as the game progresses, we generally see that we can place increasingly restrictive upper bounds on the risk aversion coefficient $\gamma$.  This can be done whenever we observe a contestant choose the action `No Deal'.  However, since we cannot place a lower bound on this parameter $\gamma$ until we observe a contestant choose the action `Deal', we cannot necessarily infer that the risk aversion level of a player is decreasing.  This is because at the point that the action `Deal' is taken, the game immediately ends.  Therefore, we cannot observe a future action choice which requires a lower $\gamma$ parameter.

We now use the examples of the real German contestant Susanne, as well as the Dutch contestant Frank (both from Post) to illustrate our concept.

\subsection{Example - Suzanne from Germany}

We start by analyzing Susanne's four remaining prizes $\left\{0.5,1\mathord{,}000,100\mathord{,}000,150\mathord{,}000\right\}$ at in round 7.  
We let $ t=7$ and work backwards from the final stage, time $t+2=9$, where two prizes remain.  Hence, we evaluate all possible outcomes from the current state at time $t$ in order to determine $Q$-values for this current state.

First, we consider the expected banker's offers for all of the possible outcomes at $t+2$.  Since for Susanne, we can only observe the Banker's offer for a single outcome, we use the observed percent of expected value of the given offer as a multiplier on the expected value of all other potential outcomes.  For time $t+2$, this is $m_{t+2}=100\%$.  Using this, we calculate the banker's function for all other possible outcomes:
\begin{align}
B(s^*_{t+2,1}&=\{0.5,1\mathord{,}000\})=m_{t+2}*\bar{s}^*_{t+2,1}=500.25\nonumber\\
B(s^*_{t+2,2}&=\{0.5,100\mathord{,}000\})=m_{t+2}*\bar{s}^*_{t+2,2}=50\mathord{,}000.25\nonumber\\
B(s^*_{t+2,3}&=\{0.5,150\mathord{,}000\})=m_{t+2}*\bar{s}^*_{t+2,3}=75\mathord{,}000.25\nonumber\\
B(s^*_{t+2,4}&=\{1\mathord{,}000,100\mathord{,}000\})=m_{t+2}*\bar{s}^*_{t+2,4}=50\mathord{,}500\nonumber\\
B(s^*_{t+2,5}&=\{1\mathord{,}000,150\mathord{,}000\})=m_{t+2}*\bar{s}^*_{t+2,5}=75\mathord{,}500\nonumber\\
B(s^*_{t+2,6}&=\{100\mathord{,}000,150\mathord{,}000\})=m_{t+2}*\bar{s}^*_{t+2,6}=125\mathord{,}000\nonumber
\end{align}
Therefore, we get the following $Q$-values for the decisions $a_{t+2,i}=0, i\in\{1,\dots ,6\}$:
namely $ Q_{t+2} \left ( s_{t+2,j}^\star \right ) $ for any $ j\in\{1,\dots ,6\}$, 
by plugging into the power utility function.

Similarly, we can work out the $Q$-values for the decisions $a_{t+2,i}=1, i\in\{1,\dots ,6\}$.  For simplicity of notation we define $Q_t(i,j)=Q_t(s^*_{t,i},a_{t,i}=j)$.
\begin{align}
Q_{t+2}(1,1)&=\frac{1}{2}\left(\frac{0.5^{1-\gamma}-1}{1-\gamma}+\frac{1\mathord{,}000^{1-\gamma}-1}{1-\gamma}\right)\nonumber\\
Q_{t+2}(2,1)&=\frac{1}{2}\left(\frac{0.5^{1-\gamma}-1}{1-\gamma}+\frac{100\mathord{,}000^{1-\gamma}-1}{1-\gamma}\right)\nonumber\\
Q_{t+2}(3,1)&=\frac{1}{2}\left(\frac{0.5^{1-\gamma}-1}{1-\gamma}+\frac{150\mathord{,}000^{1-\gamma}-1}{1-\gamma}\right)\nonumber\\
Q_{t+2}(4,1)&=\frac{1}{2}\left(\frac{1\mathord{,}000^{1-\gamma}-1}{1-\gamma}+\frac{100\mathord{,}000^{1-\gamma}-1}{1-\gamma}\right)\nonumber\\
Q_{t+2}(5,1)&=\frac{1}{2}\left(\frac{1\mathord{,}000^{1-\gamma}-1}{1-\gamma}+\frac{150\mathord{,}000^{1-\gamma}-1}{1-\gamma}\right)\nonumber\\
Q_{t+2}(6,1)&=\frac{1}{2}\left(\frac{100\mathord{,}000^{1-\gamma}-1}{1-\gamma}+\frac{150\mathord{,}000^{1-\gamma}-1}{1-\gamma}\right)\nonumber
\end{align}
Comparing these $Q$-values at $t+2$, we can determine the optimal actions at each of the possible states.  That is, choose $a_{t+2,i}=0$, $i\in\{1,\dots ,6\}$ since 
$$
Q_{t+2}(i,0)>Q_{t+2}(i,1)
$$ 
Now, we consider the expected banker's offers at each of the possible states at time $t+1$.  Here we use an expected value multiplier of $m_{t+1}=90\%$:
\begin{align}
B(s^*_{t+1,1}&=\{0.5,1\mathord{,}000,100\mathord{,}000\})=m_{t+1}*\bar{s}^*_{t+1,1}=30\mathord{,}300.2\nonumber\\
B(s^*_{t+1,2}&=\{0.5,1\mathord{,}000,150\mathord{,}000\})=m_{t+1}*\bar{s}^*_{t+1,2}=45\mathord{,}300.2\nonumber\\
B(s^*_{t+1,3}&=\{0.5,100\mathord{,}000,150\mathord{,}000\})=m_{t+1}*\bar{s}^*_{t+1,3}=75\mathord{,}000.2\nonumber\\
B(s^*_{t+1,4}&=\{1\mathord{,}000,100\mathord{,}000,150\mathord{,}000\})=m_{t+1}*\bar{s}^*_{t+1,4}=75\mathord{,}300\nonumber
\end{align}
Using these, we get the following $Q$-values for the decisions $a_{t+1,i}=0$ for $ i\in\{1,\dots ,4\}$:
\begin{align}
Q_{t+1}(1,0)&=\frac{30\mathord{,}300.2^{1-\gamma}-1}{1-\gamma} \; , \;
Q_{t+1}(2,0)=\frac{45\mathord{,}300.2^{1-\gamma}-1}{1-\gamma}\nonumber\\
Q_{t+1}(3,0)&=\frac{75\mathord{,}000.2^{1-\gamma}-1}{1-\gamma} \; , \;
Q_{t+1}(4,0)=\frac{75\mathord{,}300^{1-\gamma}-1}{1-\gamma}\nonumber
\end{align}
Using the optimal actions at $t+2$, we can calculate the continuation $Q$-values at $t+1$, that is where $a_{t+1,i}=1, i\in\{1,\dots ,4\}$:
\begin{align}
Q_{t+1}(1,1)&=\frac{1}{3}\left(\frac{500.25^{1-\gamma}-1}{1-\gamma}+\frac{50\mathord{,}000.25^{1-\gamma}-1}{1-\gamma}+\frac{50\mathord{,}500^{1-\gamma}-1}{1-\gamma}\right)\nonumber\\
Q_{t+1}(2,1)&=\frac{1}{3}\left(\frac{500.25^{1-\gamma}-1}{1-\gamma}+\frac{75\mathord{,}000.25^{1-\gamma}-1}{1-\gamma}+\frac{75\mathord{,}500^{1-\gamma}-1}{1-\gamma}\right)\nonumber\\
Q_{t+1}(3,1)&=\frac{1}{3}\left(\frac{50\mathord{,}000.25^{1-\gamma}-1}{1-\gamma}+\frac{75\mathord{,}000.25^{1-\gamma}-1}{1-\gamma}+\frac{125\mathord{,}000^{1-\gamma}-1}{1-\gamma}\right)\nonumber\\
Q_{t+1}(4,1)&=\frac{1}{3}\left(\frac{50\mathord{,}500^{1-\gamma}-1}{1-\gamma}+\frac{75\mathord{,}500^{1-\gamma}-1}{1-\gamma}+\frac{125\mathord{,}000^{1-\gamma}-1}{1-\gamma}\right)\nonumber
\end{align}
Thus, we can see that:
\begin{align}
Q_{t+1}(1,0)&<Q_{t+1}(1,1) \text{ if } \gamma<0.22617 \; , \; Q_{t+1}(2,0) <Q_{t+1}(2,1) \text{ if } \gamma<0.22077\nonumber\\
Q_{t+1}(3,0)&<Q_{t+1}(3,1) \text{ if } \gamma<1.50645 \; , \; Q_{t+1}(4,0)<Q_{t+1}(4,1) \text{ if } \gamma<1.54085\nonumber
\end{align}
Since we observe that Suzanne, when faced with $s^*_{t+1,4}$, chose action $a_{t+1,4}=1$, we must have that her $\gamma<1.54085$.

Finally, we can now consider the banker's offer at the current state at time $t$.  Here, we see that the expected value multiplier is $m_t=73.31\%$:
$$
B(s^*_t=\{0.5,1\mathord{,}000,100\mathord{,}000,150\mathord{,}000\})=m_t*\bar{s}^*_{t}=46\mathord{,}000
$$
Therefore, we get the following $Q$-value for the decision $a_t=0$:
$$
Q_t(s^*_t,a_t=0)=\frac{46\mathord{,}000^{1-\gamma}-1}{1-\gamma}
$$
Using the optimal actions at $t+1$, we calculate the continuation $Q$-values at $t$, with $a_t=1$
and determine
$$
Q_t(s^*_t,a_t=1)=\left\{
\begin{array}{rl}
\frac{1}{4}\left[Q_{t+1}(1,1)+Q_{t+1}(2,1)+Q_{t+1}(3,1)+Q_{t+1}(4,1)\right]&\text{if }\gamma<0.22077\\
\frac{1}{4}\left[Q_{t+1}(1,1)+Q_{t+1}(2,0)+Q_{t+1}(3,1)+Q_{t+1}(4,1)\right]&\text{if }0.22077<\gamma<0.22617\\
\frac{1}{4}\left[Q_{t+1}(1,0)+Q_{t+1}(2,0)+Q_{t+1}(3,1)+Q_{t+1}(4,1)\right]&\text{if }0.22617<\gamma<1.50645\\
\frac{1}{4}\left[Q_{t+1}(1,0)+Q_{t+1}(2,0)+Q_{t+1}(3,0)+Q_{t+1}(4,1)\right]&\text{if }1.50645<\gamma<1.54085
\end{array}
\right.
$$
where $Q_{t}(i,j)=Q_{t}(s^\star_{t,i},a_{t,i}=j)$.

Using our optimal choice of $a_{t+1}$ for a given value of $\gamma$ we have:
\begin{itemize}
\item If $\gamma<0.22077$, then $Q_t(s^*_t,a_t=0)<Q_t(s^*_t,a_t=1)$ for $\gamma<0.68666$.
\item If $0.22077<\gamma<0.22617$, then $Q_t(s^*_t,a_t=0)<Q_t(s^*_t,a_t=1)$ for $\gamma<0.87506$.
\item If $0.22617<\gamma<1.50645$, then $Q_t(s^*_t,a_t=0)<Q_t(s^*_t,a_t=1)$ for $\gamma<2.61618$.
\item If $1.50645<\gamma<1.54085$, then $Q_t(s^*_t,a_t=0)<Q_t(s^*_t,a_t=1)$ for $\gamma<2.73117$.
\end{itemize}
Therefore, for $\gamma<1.54085$ the optimal decision at time $t$ is $a_t=1$, `No Deal', since the continuation value is the larger.  This is consistent with the choice that Susanne made.

\begin{figure}[htb]
\centering
\includegraphics[width=1\textwidth]{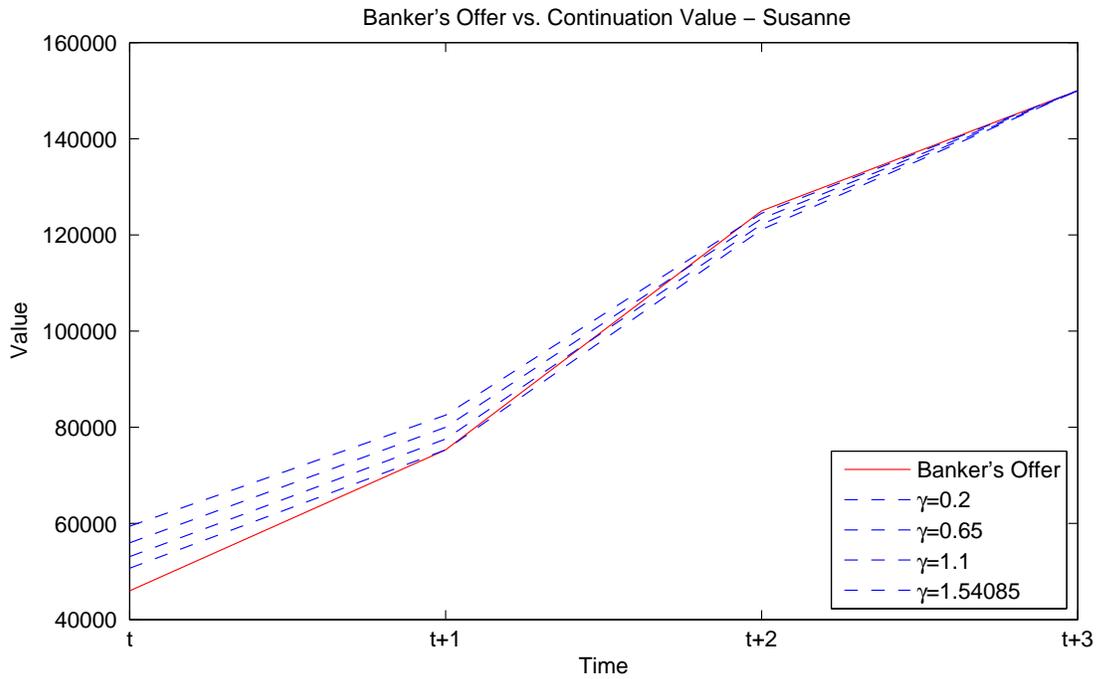}
\caption{\textsl{\small{Susanne's evolving values of the two actions for varying levels of risk aversion}}}
\label{susanne}
\end{figure}

Now, we can plot the evolving monetary values of each action $a$ at each point in time. In figure \ref{susanne}, the solid line represents the value of the Banker's offer ($a=0$), and the four dotted lines represent the continuation value ($a=1$) for a range of ascending values of the risk aversion parameter $\gamma$.  At each point in time, the contestant will choose the action with the larger value.  Therefore, we see that Susanne will prefer action $a_{t+1}=1$ at time $t+1$ if her risk aversion parameter is $\gamma<1.54085$.  However, at time $t+2$, all values of $\gamma>0$ will induce the optimal action $a_{t+2}=0$ since the value of the Banker's offer is larger than all possible continuation values.  Naturally, at time $t+3$, where the contestant holds only one case, there is complete certainty about the prize value, therefore (for completeness) the Banker's offer has converged to the value of that prize.

Also, it is important to note that although we have only considered the final stages of Susanne's game, it is unlikely that an earlier stage has placed a more restrictive upper bound on the risk aversion parameter.  Since the Banker's offer as a percentage of expected value is increasing over time, choosing the action `Deal' early on in the game becomes very unattractive.  Only a contestant with a very large risk aversion level will choose `Deal' early and forgo the almost certain increasing percentage of expected value implicit in the Banker's offer function.

Thus, we can see that given a constant risk aversion level of $\gamma<1.54085$, it is possible to rationally observe the actions of the real player, Susanne, except for the final time period $t+2$.  Seemingly, only a completely risk-neutral or even risk-seeking player would be willing to turn down an expected value offer in exchange for a fair gamble on a set of prizes as per Jensen's Inequality.  However, we have observed Susanne turning down an offer of $125\mathord{,}000$ for a $50/50$ gamble on the set of prizes $\{100\mathord{,}000,150\mathord{,}000\}$.  What could cause her to do so while allowing us to not deviate from the constant risk aversion level hypothesis?

\subsection{Example - Frank from the Netherlands}
Consider Frank's four remaining prizes $\{0.50,10,20,10\mathord{,}000\}$ in round $t=7$.  As he plays through his remaining rounds, Frank is presented with the following remaining prizes at each period:
$$
s_t=\{0.5,10,20,10\mathord{,}000\}, s_{t+1}=\{10,20,10\mathord{,}000\}, s_{t+2}=\{10,10\mathord{,}000\},s_{t+3}=\{10\}
$$
At each one of these periods, we observe the following Banker's offers and implied $m$ values:
\begin{align}
B(s_t)=2\mathord{,}400=m_t*\bar{s}_t=m_t*2507.63\quad&\Rightarrow\quad m_t=95.71\%\nonumber\\
B(s_{t+1})=3\mathord{,}500=m_{t+1}*\bar{s}_{t+1}=m_{t+1}*3343.33\quad&\Rightarrow\quad m_{t+1}=104.69\%\nonumber\\
B(s_{t+2})=6\mathord{,}000=m_{t+2}*\bar{s}_{t+2}=m_{t+2}*5005.00\quad&\Rightarrow\quad m_{t+2}=119.88\%\nonumber
\end{align}
Interestingly, the Banker's offer percentage is very high with respect to expected value.  In fact, the final two offers are significantly above expected value.

At time period $t+2$ we can see that Frank turns down the offer of $6\mathord{,}000$ for the 50/50 gamble between $10$ and $10\mathord{,}000$.  Short of Frank being risk averse and acting completely irrationally, there are two possible explanations for this observed action.  Either he truly has a risk aversion level characterized by $\gamma<0$ (i.e. he is risk seeking), or he has a large ``enjoyment'' benefit from playing the game (i.e. his $b$ is quite large).  It is possible that if he has a large enough value of $b$, that his choice of `No Deal' at time $t+2$ is not risk seeking, but actually risk averse!

\begin{figure}[htb]
\centering
\includegraphics[width=1\textwidth]{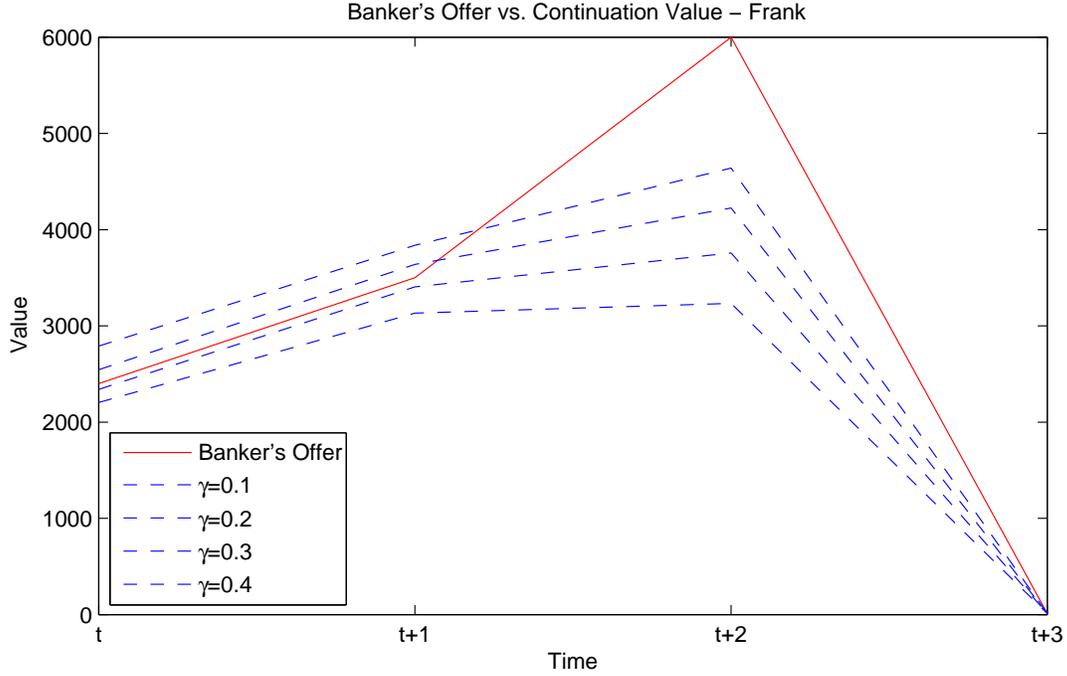}
\caption{\textsl{\small{Frank's evolving values of the two actions for varying levels of risk aversion}}}
\label{frank}
\end{figure}

As an additional point, looking at his decision at time $t+1$, we can see that even though Frank turns down a Banker offer larger than the expected value, this choice may \emph{not} have been risk seeking even if he has an ``enjoyment'' benefit of $b=0$!  To see this, consider the continuation value: if Frank expects that the Banker's expected value offer percentage $m$ is increasing in each round, he may get an even larger Banker's expected value offer percentage $m$ in the next round.  Therefore, even if the $m$ value in one round is larger than $100\%$, an observed action of `No Deal' does not preclude risk aversion if an even larger $m$ value is expected in the next period.  Plainly, this is caused by a large continuation value.

Similarly as we did for Susanne, figure \ref{frank} shows the evolving 
monetary values of each action $a$ at every point in time.

\subsection{The terminal risk-seeking choice}

One interesting feature of Suzanne's and Frank's choices are their risk seeking behaviour at the terminal decision.
Clearly, there is no continuation value left in the game. How irrational are these decisions?
They are gaining enjoyment from other sources such as: enthusiasm from playing the game, audience encouragement, and even excitement from being on TV.  
If we assume that the marginal enjoyment benefit is positive at each further stage of the game since a player would likely prefer to play the game longer than shorter.
If we let the marginal benefit to Susanne to play the final round of the game to be the value $b$, 
we then have the following condition to turn down the Banker's final offer:
$$
\frac{125\mathord{,}000^{1-\gamma}-1}{1-\gamma}\le\frac{1}{2}\left[\frac{(100\mathord{,}000+b)^{1-\gamma}-1}{1-\gamma}+\frac{(150\mathord{,}000+b)^{1-\gamma}-1}{1-\gamma}\right]
$$
$$
\gamma=1.54085\quad\Rightarrow\quad b=3\mathord{,}761.90
$$
Therefore, we can see that even at the highest possible risk aversion level of $\gamma=1.54085$, Susanne only needs an ``enjoyment'' benefit of \euro $3\mathord{,}761.90$ in order to justify her `No Deal' choice.
\section{Discussion}

The TV game show \textit{Deal or No Deal} is well suited for analyzing risky choices. The stakes are high and the outcomes for the contestants can range from a multi-mullion payday to empty-handed.  The game involves only binary decisions and there are no subjective probabilities as the odds are well-defined ahead of time. We show, however, that the sequential nature of the game induces a significant continuation value that has been missed by a number of researchers.  Choices can appear irrational and risky when in fact they are just an optimal strategy with reasonable (constant) parameters for risk aversion in a simple CRRA model of utility and choice.

Our approach uses $Q$-learning, a popular technique in the reinforcement learning literature, to solve for the optimal strategy. We then show that choices provide bounds on risk aversion parameters conditional on optimal play. We analyze the Frank and Suzanne data-sets from Post et al (2008) which tests the Thaler and Johnson (1990) hypothesis that many choices are affected by past outcomes. While Post et al (2008) conclude that many of the choices are the effect of time-varying risk aversion, our empirical findings are more in line with Bombardini and Trebbi (2007) who propose a dynamic expected utility model for the Italian version of the game and conclude that contestants have constant levels of risk aversion but players are extremely heterogeneous in their beliefs.

We note that the idea of time-varying risk aversion has been captured in the economics and finance literature in a number of ways. A common approach is via habit formation as shown by Campbell and Cochrane (1999) and Brandt and Wang (2011).  Nevertheless, this seems unrealistic in such a short lived game. Other interesting utility functions -- where Q-learning can easily be applied -- are prospect theory which predicts possible time inconsistencies in gambling attitudes as discussed by Barberis (2011) or recursive utility methods as proposed in Kreps and Porteus (1978). We emphasize here that the neo-classical constant relative risk aversion model provides a realistic model for the contestants' attitudes studied here except for their final risk seeking choices.

\section{References}

\noindent Banks, D., F. Petralia and S. Wang (2011). Adversarial Risk Analysis: Borel Games.
\emph{Applied Stochastic Models}, 27(2), 72-86.\medskip

\noindent Barberis, N. (2011). A Model of Casino Gambling. \textit{Working Paper}.\medskip

\noindent Barberis, N., M. Huang and T. Santos (2001). Prospect Theory and Asset Prices.
\textit{The Quarterly Journal of Economics}, 116(1), 1-53.\medskip

\noindent{B}ellman, R. (1957). \textit{Dynamic Programming}. Princeton
University Press.\medskip

\noindent{B}ertsekas, D. (1994) \textit{Dynamic Programming and Optimal
Control} (vols I and II). Athena Scientific, Belmont, MA.\medskip

\noindent{B}ertsekas, D. and Tsitsiklis, J.N. (1996) \textit{Neuro-Dynamic
Programming}. Athena Scientific, Belmont, MA.\medskip

\noindent Bombardini, M. and F. Trebbi (2007). Risk aversion and Expected Utility Theory:
an Experiment with large and small stakes. \emph{Working Paper}.\medskip

\noindent Brandt, M. and L. Wang (2010). Measuring the Time-Varying Risk-Return Relation from the Cross-Section of Equity. \emph{Working Paper}.\medskip

\noindent Camerer, C. (2003). Behavioral Game Theory: Experiments in Strategic Interaction.
\textit{Princeton University Press}.\medskip

\noindent Campbell J. and J. Cochrane (1999). By Force of Habit: A Consumption-Based Explanation of Aggregate Stock Market Behavior. \textit{The Journal of Political Economy}, 107(2), 205-251.\medskip

\noindent de Finetti, B. (1952). Sulla preferibilit\'a. \emph{Annali di Economia}, 11, 658-709.\medskip

\noindent de Finetti, B. (1974). \textit{The Theory of Probability},
Vol I,II Wiley, Chichester.\medskip

\noindent Gittins, J.C. (1979). Bandit processes and dynamic allocation
indicies (with discussion). \textit{Journal of Royal Statistical Society B},
41, 148-77.\medskip

\noindent Holt, C.A. and S.K. Laury (2002). Risk Aversion and Incentive Effects.
\emph{American Economic Review}, 92(5), 1644-55.\medskip

\noindent{H}oward, R. (1960). \textit{Dynamic Programming and Markov
Processes}. MIT Press, Cambridge.\medskip

\noindent Kelly, J. (1956). A new interpretation of Information rate.
\textit{Bell System Technical Journal}, 35, 917-926.\medskip

\noindent Kreps, D. and E. Porteus (1978). Temporal Resolution of Uncertainty and Dynamic Choice Theory. \textit{Econometrica}, 46(1), 185-200.\medskip

\noindent Nau, R. (2011). Risk, Ambiguity and State-Preference Theory.
\emph{Economic Theory}, forthcoming.\medskip

\noindent Polson, N.G. and M. Sorensen (2011). A Simulation-Based Approach to Stochastic Dynamic Programming.
\emph{Applied Stochastic Models}, 27(2), 151-163.\medskip

\noindent Post, T., N.J. van der Assem, G. Baltussen and R. H. Thaler (2008).
Deal or No Deal? Decision making under risk in a large-payoff game show.
\textit{American Economic Review}, 98(1), 38-71.\medskip

\noindent{P}uterman, M. (1994). \textit{Markov Decision Processes}. Wiley: New York.\medskip

\noindent Ramsey, F.P. (1931). Truth and Probability, In \emph{The
Foundations of Mathematics and other Logical Essays.} 156-198,
Routledge and Kegan Paul, London.\medskip

\noindent{S}utton, R. and Barto, A.G. (1998). \textit{Reinforcement Learning}. 
MIT Press, Cambridge.\medskip

\noindent Thaler, R.H. and E.J. Johnson (1990). Gambling with the House Money and trying to Break Even:
the effects of prior outcomes on risky choice. \emph{Management Science}, 36(6), 643-660.\medskip

\noindent von Neumann, J. and O. Morgenstern (1944). \emph{Theory of Games and Economic Behaviour}.
Princeton University Press.\medskip

\noindent Watkins, C. (1989). Learning from Delayed Rewards. \textit{PhD. Thesis, Cambridge University}.\medskip

\noindent Watkins, C. and Dayan (1992). Q-Learning. \textit{Machine Learning}, 8, 279-292.\medskip

\noindent{W}hittle, P. (1983). \textit{Optimization over Time: Vols I and II.} 
Wiley: New York.\medskip

\begin{table}
\caption{Susanne's Choices}
\vspace{0.2in}
\footnotesize
\begin{tabular}{cccccccccc}
\hline\hline
Prize         & 1 & 2 & 3 & 4 & 5 & 6 & 7 & 8 & 9 \\\hline
\euro 0.01    & $\times$ & $\times$ & $\times$ & $\times$ & & & & & \\
\euro 0.20    & $\times$ & $\times$ & $\times$ & & & & & & \\
\euro 0.50    &  $\times$ & $\times$ & $\times$ & $\times$ & $\times$ & $\times$ & $\times$ & & \\
\euro   1     & & & & & & & & &\\
\euro   5     & & & & & & & & & \\
\euro   10    & & & & & & & & &\\
\euro   20    & $\times$ & $\times$ & & & & & & & \\
\euro   50    & $\times$ & $\times$ & & & & & & & \\
\euro  100    & $\times$ & $\times$ & $\times$ & $\times$ & & & & & \\
\euro  200    & & & & & & & & & \\
\euro  300    & $\times$ & $\times$ & $\times$ & & & & & & \\
\euro  400    & $\times$ & & & & & & & & \\ 
\euro  500    & & & & & & & & & \\
\euro 1,000   & $\times$ & $\times$ & $\times$ & $\times$ & $\times$ & $\times$ & $\times$ & $\times$ & \\
\euro 2,500   & $\times$ & $\times$ & $\times$ & $\times$ & $\times$ & $\times$ & & & \\
\euro 5,000  & $\times$ & & & & & & & & \\
\euro 7,500  & & & & & & & & & \\
\euro 10,000 & $\times$ & $\times$ & & & & & & & \\
\euro 12,500 & $\times$ & $\times$ & $\times$ & & & & & & \\
\euro 15,000 & $\times$ & & & & & & & & \\
\euro 20,000 & $\times$ & $\times$ & & & & & & & \\
\euro 25,000 & $\times$ & $\times$ & $\times$ & $\times$ & $\times$ & & & & \\
\euro 50,000 & $\times$ & & & & & & & & \\
\euro 100,000 & $\times$ & $\times$ & $\times$ & $\times$ & $\times$ & $\times$ & $\times$ & $\times$ & $\times$\\
\euro 150,000 & $\times$ & $\times$ & $\times$ & $\times$ & $\times$ & $\times$ & $\times$ & $\times$ & $\times$\\
\euro 250,000 &  $\times$ &  &   &  & & & & & \\ \hline
Average \euro & 32,094 & 21,431 & 26,491 & 34,825 & 46,417 & 50,700 & 62,750 & 83,667 & 125,000\\
Offer \euro & 3,400 & 4,350 & 10,000 & 15,600 & 25,000 & 31,400 & 46,000 & 75,300 & 125,000\\
Offer \% & 11\% & 20\% & 38\% & 45\% & 54\% & 62\% & 73\% & 90\% & 100\%\\
Decision & No Deal & No Deal & No Deal & No Deal & No Deal & No Deal & No Deal & No Deal & No Deal \\\hline
\end{tabular}
\normalsize
\end{table}

\begin{table}
\caption{Frank's Choices}
\vspace{0.2in}
\footnotesize
\begin{tabular}{cccccccccc}
\hline\hline
Prize         & 1 & 2 & 3 & 4 & 5 & 6 & 7 & 8 & 9 \\\hline
\euro 0.01    & $\times$ & $\times$ & & & & & & & \\
\euro 0.20    & $\times$ & $\times$ & & & & & & & \\
\euro 0.50    & $\times$ & $\times$ & $\times$ & $\times$ & $\times$ & $\times$ & $\times$ & & \\
\euro   1     & $\times$ & $\times$ & $\times$ & $\times$ & $\times$ & & & &\\
\euro   5     & & & & & & & & & \\
\euro   10    & $\times$ & $\times$ & $\times$ & $\times$ & $\times$ & $\times$ & $\times$ & $\times$ & $\times$\\
\euro   20    & $\times$ & $\times$ & $\times$ & $\times$ & $\times$ & $\times$ & $\times$ & $\times$ & \\
\euro   50    & & & & & & & & & \\
\euro  100    & & & & & & & & & \\
\euro  500    & & & & & & & & & \\
\euro 1,000   & $\times$ & & & & & & & & \\
\euro 2,500   & $\times$ & $\times$ & $\times$ & & & & & & \\
\euro 5,000  & $\times$ & $\times$ & & & & & & & \\
\euro 7,500  & & & & & & & & & \\
\euro 10,000 & $\times$ & $\times$ & $\times$ & $\times$ & $\times$ & $\times$ & $\times$ & $\times$ & $\times$\\
\euro 25,000 & $\times$ & $\times$ & & & & & & & \\
\euro 50,000 & $\times$ & $\times$ & $\times$ & $\times$ & & & & & \\
\euro 75,000 & $\times$ & $\times$ & $\times$ & & & & & & \\
\euro 100,000 & $\times$ & $\times$ & $\times$ & & & & & &\\
\euro 200,000 & $\times$ & $\times$ & $\times$ & $\times$ & & & & & \\
\euro 300,000 & $\times$ & & & & & & & & \\
\euro 400,000 & $\times$ & & & & & & & & \\
\euro 500,000 & $\times$ & $\times$ & $\times$ & $\times$ & $\times$ & $\times$ & & & \\
\euro 1,000,000 & $\times$ & & & & & & & & \\
\euro 2,500,000 & & & & & & & & & \\
\euro 5,000,000 & $\times$ & & & & & & & & \\ \hline
Average \euro & 383,427 & 64,502 & 85,230 & 95,004 & 85,005 & 102,006 & 2,508 & 3,343 & 5,005\\
Offer \euro & 17,000 & 8,000 & 23,000 & 44,000 & 52,000 & 75,000 & 2,400 & 3,500 & 6,000\\
Offer \% & 4\% & 12\% & 27\% & 46\% & 61\% & 74\% & 96\% & 105\% & 120\%\\
Decision & No Deal & No Deal & No Deal & No Deal & No Deal & No Deal & No Deal & No Deal & No Deal \\\hline
\end{tabular}
\normalsize
\end{table}

\end{document}